# An Open Data and Geo-based Information Systems

[1]Dexter I. Mercurio, [2]Alexander A. Hernandez

Technological Institute of the Philippines, Manila, Philippines
Email: [1]dex_mercurio@yahoo.com.ph, [2]alexander.hernandez@tip.edu.ph
Contact: [1]+639073818003, [2]+639154860445

**Abstract:** Barangay is the smallest type of government in the Philippines, and it is driven and represented by its barangay authorities. The barangay officials are accountable for keeping the records of citizens health and crime incidents. It also the first-hand source of information of the national government to develop government programs, community services, and maintain peace and order. This paper presents a developed a web-based information system incorporating open data and geo-based features for a pilot community in the Philippines. This system serves as a platform for information collection and used for planning, analysis, decision-making and increase effectiveness and efficiency of government services in the community.

## I. INTRODUCTION

The quality of service of the government authorities is an essential element for the development of a community. In the Philippines Barangay first-hand source of information and services is the barangay. The barangay council is lead by the Chairman and seven barangay councilor, youth council chairperson, barangay secretary and Treasurer. They are intended to practice conventional and administrative forces. They are in charge of keeping the records of everyone living in the community, and all these files are valuable for different purposes. The barangay is responsible of disseminating information such as government programs, medical and health missions, and the first line of safeguard during emergency, disaster preparedness, and security for residents.

The absence of information systems in small communities in the Philippines, early warning information, and dissemination, lack of training and information about what to do when a calamity happens, absence of coordination among authorities in the government and community with regards to hazard warning is a complex problem that needs to be addressed in most local communities [1]. The barangay level is considered as a critical partner of the national government in local development undertakings in the country. It is the primary partner of the national government in economic advancement efforts in the country. It is because of their key strategic situations being at the grassroots level where competent and proficient administrations are given to roles and responsibilities to handle a community of residents [2]. Barangay should deliver efficient and effective services to the residents. Thus, providing community information can enhance decision-making, planning, and access to the real situation and improvement activities of community health, information ought to shape the foundation of decision and the impact of change in all areas [3].

In spite of the expanding need, however, little work has been done in the generation of data, and the audacity in planning is minimized because of the shortage and limitation of information collected from small communities such as the barangay [4]. These situation typically happens in the Philippines in which most barangays have no information systems to collect, store, analyze and use data for planning, development of government health programs, emergency and disaster preparedness, monitor crime and enhance safety and security of the residents. Also, the national government have limited access to barangays information on profile of residents, health and crime information, which hinders the development of sufficient government support for local residents in the community.

Hence, to address these problems, this study aims to develop a community information systems integrating open data and geo-based mapping features. The system aims to collect information on the profile of residents, monitor health and crime records, as well as provide real-time access to reports for the national government.

## II. Related Literature

"The Local Government Code of 1991 (Republic Act 7160) has the power, authority, and duty from the central government to Local Government Units (LGUs) in performing essential legislative functions"[5]. Among LGUs obligations are to undertake development planning. The National Statistics Office has been vigorous depended upon by LGUs to give them the information that they require especially in the statistic sector. Because of budgetary limitations, the workplace couldn't generally give every single LGU what they need.LGUs need to rely on upon their Socio-Economic and Physical Profile (SEPP). The database contains information and statistic on the social, economic and physical situation, and other fundamental data about the LGU. The database in this manner ought to be exact, reliable and timely [5].

Automation has proven its value in various agencies of the Philippine government. Since 1971 computerization in the Philippine government was already existent when the National Computer Center NCC) was established by Executive Order 322. From that, the government gave policies and initiatives on how to enhance and protect technology in the Philippines. Government Information Systems Plan (GISP) as a structure and guide or all computerization effort in the government (EO 265). One of the objectives of GISP is to make a system of governance that will lead to quicker and better delivery of public goods and services. The utilization of technology in government is practiced in the most progressive nations only. The Philippines, characterized by the World Bank as a nation with lower middle-income level, still has a potential for e-government [6].

Statistical information tool to acquaint people about the quantitative parts of their community life and improvement. On the off chance that no data will be accessible as the basis for planning at the local level, services required and development goals will never be realized [4]. That is the reason a barangay information framework that can examine raw information into statistical data is an incredible help for the barangay authority and staff and government offices to minimize their work.

In recent studies show that community information system is turning into an engaging force and undeniably fundamental approach to distributing administrative data from local governments. These frameworks are utilizing administrative data to make and convey vital community and a social pointer to concerned people, social management and community improvement [7,8]. It is about upgrading community level in administration and services utilizing cutting edge communication tools, and information it is a crucial piece of community management [9].

In the Philippines, barangay is also in charge of monitoring the community peace and order and health status. However the health information system of the Philippines suffers experienced a few issues: Lack of quality in information gathering, Data repetition, Missing or inadequate report generation and compilation tools [10]. Because of this, the government reaction for medical services usually in rural communities is delayed [11].

In some recent studies expressed that community information system can be utilized as an instrument for gathering health information of the ccommunity that can be utilized as a part of surveying the present community health status and assess the adequacy of health services for community information system is a viable tools or can give guidelines in decisions making and planning in their Health community program [12, 13, 14].

Another problem of the barangay official's face is how to inform and educate the barangays residents about government programs. Immediate dissemination of information is still issued in remote areas.

Open data is vital in promoting well inform public. It is a key in research and door to a more extensive knowledge and innovation [15, 16]. However, data or information subject to privacy should be kept private [17]. Due to budgetary constraints, the government could not provide each and every LGU an Information System.

The Growing accessibility of Information technologies has empowered law enforcement agencies to gather detailed information about different crimes. Classification techniques can be applied to this information to manufacture decision guide tools for law enforcement [18]. Decision Trees are thought to be a considered amongst the most popular methodologies for representing classifiers. Researchers from different disciplines, for example, pattern, statistic, machine learning. A decision tree is a classifier communicated as a recursive partition of the instance space. The primary edge of utilizing decision tree is that it is easy to comprehend and translate [19]. The goal of mining data is to discover patterns as a tool for translating the data and make predictions about it. Decision trees are one approach for such forecasts [20].

Crime is essentially "unpredictable" incident. Space and time do not compel it. It completely relies on upon human behavior. Different activities performed by criminal create a large volume of data, and again this data can be available in a range of formats. Extricating all available unusual numerous patterns based on the crime factors and applied classification approach to anticipate the likelihood of a man to commit a crime once more. An answer for the criminal forecasting issue utilizing naive bayes theory. The criminal forecast issue reaches as finding likelihood criminal of a specific crime incident when the history of crime incidents is given with incident-level crime data. The system is proved for the criminal forecasting issue utilizing the cross-validation, and the test comes about demonstrate that the proposed framework provides high scores in a finding of suspected criminals [21].

### III. Methodology

This project uses Agile Software Development method to ensure that all user specification, development, and outcomes are meet. The study uses PHP, JQUERY, JAVASCRIPT, AJAX programming and third party technologies Google maps and Itexmo SMS API. The study includes different users including barangay secretary and treasurer, health workers, local government agencies, and residents. This research used a software evaluation following ISO9126 criteria and distributed to the users during the testing stage. The Likert's scale with the interpretation of High Acceptable, Moderately Acceptable, Acceptable, Slightly Acceptable, and Not Acceptable was used to specify the users' level of agreement or disagreement on the software evaluation items. The results of software evaluation are present in the succeeding sections of this paper.

## IV. Result and Discussion

In this section, the outcome of the study is presented. An open data and geobased information system figure 1 present the system architecture in web and mobile platform. The web module contains registration/records management, clearance, and certification management, blotter and complaints management, community and medical records, geobased and statistic report, SMS notification and open data page.

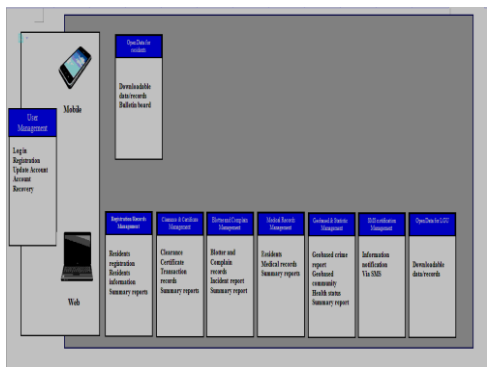

**Figure 1: An Open data and Geobased Information System**

While the mobile module includes barangay open data, these modules are beneficial for the barangay officials, health workers, and local government agencies.

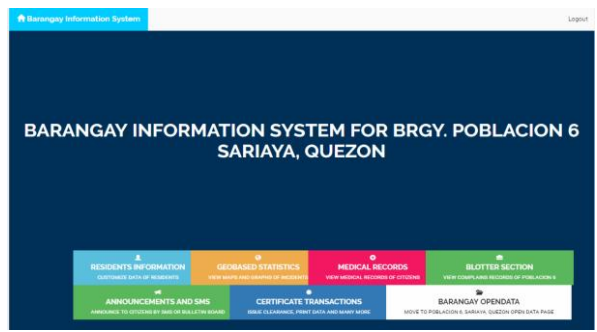

**Figure 2: An Open Data and Geobased Information System**

**Figure 3: Registration/Records Management**

Figure 3 presents the registration of the barangay residents' information, each resident is given an identification number for efficient monitoring. This resident number serves as their tracking number. Information of every resident in the barangay can be seen in this module like their transaction made with the barangay and complains record if there's any. Instead of keeping files in the cabinet it gives a safer way of keeping data. It allows the user to verify the identity of the person quickly the long procedure of locating resident data is minimized. User can give fast and efficient services.

**Figure 4: Clearance and certificate management**

Figure 4 user can issue and print a certificate for the client. It verifies first the identity of the resident then checks the record if the resident has pending or existing complaint or case record in the barangay. It also display the previous transactions with barangay officials, the number of times and the purpose of clearance requested. All transactions are in a centralized database system and added to the resident information server. Transactions done by the residents are recorded for the purpose of generating summary reports of profiles. Monitoring the residents requesting for certification and clearance is not an easy task. It is important to verify the identity of the person carefully before the treasurer issue certificate with this feature the user can quickly view the transaction history made by the residents and his or her profile. This is to help to minimize or to avoid the misuse of barangay clearance and assist the user in monitoring their barangay transaction.

**Figure 5: Blotter and Complain Management**

Crimes and complaint brought into barangay were recorded and given an incident report number. The incident, a person complained, and the complainant/s is recorded in the database systems and automatically added to the information of the persons involve in the incident. This help official to have a safe and organized crime and incident database that can quickly retrieve if needed. Having accessibility to certain information help authorities to analyze the current peace and order situation in the community, and make the immediate plan for the current crime problems, it can enhance the community policy creation, and resource management.

**Figure 6: Community health record management**

Figure 6 presents medical information of residents in the community. This data serve then as a guide or indicator of the effectiveness of their program and help them to evaluate the current population health status. Health cases are recorded and added to the residents' information; it gives the health worker easy way to access and manage their data. Based on this information health authorities can make approaches and programs how to improve health in a community, maintaining and monitor the effectiveness of their health programs and immediately response to an identified community health need. It improves community healthcare services and enhances the health of the community.

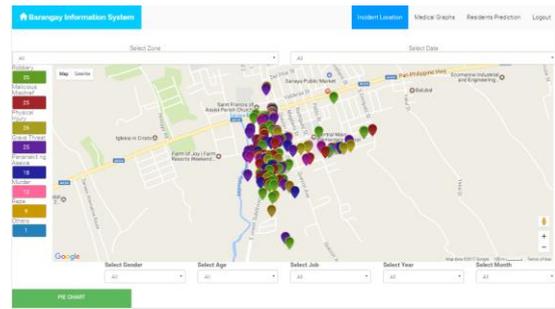

**Figure 7: Geobased Information System**

Figure 7 presents the geobased information feature. It uses third party technology Google map APIs to incorporate mapping components of crime and health information. Zone separates the barangay (community). There is 7 zones covering different streets and all parts of the barangay. The Google Maps API is the primary reference utilizing the maps features integrated with the system. Crime related incident happened in the barangay are shown on the map. Using the map, it gives the user access to locate the most crime-prone area in the barangay. Places in the barangay where an incident happened are marked in different colors for distinction. Health-related cases in the barangay are also seen on the map places and where medical attention is needed quickly as recognized through the use of Google API map.

Identifying the crime hot spot in a community is a major factor in preventing crime, Searching incident places mostly happened in the past could help authorities make better crime policies and guidelines. Authorities can prevent or reduce crime at these high crime places and provide a better safeguarding of the community. Using GIS, the area of a particular medical problem could be easily monitored and layout the historical disease information. The area in the community that needed a medical attention can easily be recognized. Using GIS, community health authorities can deliver the right services or health program for different areas of the community.

**Figure 8: SMS Notification System**

Figure 8 presents the use of SMS (short messaging service) to disseminate important announcement or advisory. The user can send a text message through SMS feature. It serves them as communication tool especially in times of calamity, emergency and disaster situations, it can offer efficient information and located evacuation center. Government programs and medical mission can advise here. With the use of third party technology, Itexmo SMS API, messages can be sent to residents' mobile number registered in the profile of community residents.

SMS is one of the commonly used communication media in getting information Its cheap, fast and efficient alternative way of spreading information to the community instead of going house to house or use flyers. Information sent through SMS saves time, effort, and reaches more target residents. It ensures that government policies, programs, and information reach every resident of the community.

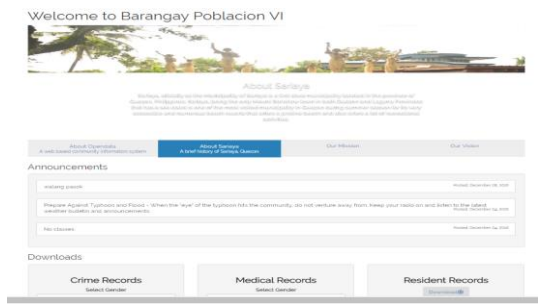

**Figure 9: Open data for residents and LGU's**

It is where data that is not subject to privacy is share for the inhabitants of the barangay and LGU's. This data and statistic can be seen and download in Excel format. Like government programs, advisories and barangay information, crime status, barangay health status. This data can help them in their researches. All data can be downloaded in Excel format.

It gives the community better insights of the government policies and the current situation of the community. They can use this data to produce new services, programs and it promotes a well-informed community.

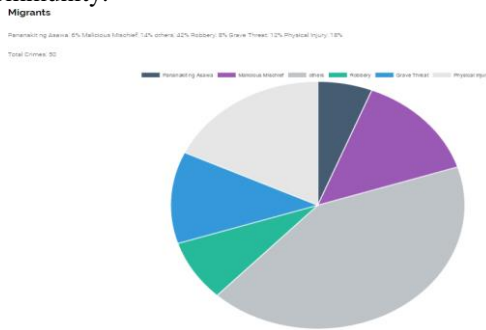

**Figure 10: Crime Chart**

Our goal is to use Decision Trees and Naive Bayesian to predict crime and evaluate the correlation of crime with the migrants in the barangay. Predictors is based on their residency, gender, age occupation,and date crime committed. , these variables are used to know if it influences whether a non-migrant or migrant do a crime. Naive Bayesian was used to the calculate the likelihood of the of the non-migrant, migrant residents to commit a crime. These predictors are used to determine the common violation of the offenders. Employment, Alcohol Problems, Family Problems, Status, Drug Problems, Age, Gambling, drug addiction, mental health Problems, Financial Problems, school problem, Gender are the factors help to determine the probabilities of each type of the residents to commit crimes.

This gives alternative way for the community authorities to evaluate the current peace and order status. They can analyze carefully the cause of crime in their community and make the appropriate action to prevent it.This can guide them in making decision in crime policing and allocating crime fighting resources efficiently. It benefits not just the user but the whole community. Reducing crime risk ,utilization limited resources and better security plan for the safety of the public.

This system gives the barangay staff and official a tool that benefits both them and community residents. Barangay authorities can provide fast and quality services, and for the decision maker as a guide for their community programs

Table 1 Summary of Software Evaluation

| Criteria | Mean | Interpretation |
|---|---|---|
| 1. Functionality | 4.5 | Acceptable |
| 2. Reliability | 4.5 | Highly Acceptable |
| 3. Usability | 4.4 | Acceptable |
| 4. Efficiency | 4.4 | Acceptable |
| 5. Maintainability | 4.4 | Highly Acceptable |
| 6. Portability | 4.6 | Acceptable |
| **Total** | **4.6** | **Acceptable** |

In summary, the software evaluation indicates a strong perception among the respondents that the Barangay information system is highly functional (4.5), reliable (4.5), usable (4.4), efficient (4.4), maintainable (4.4, and portable (4.6). Hence, the software evaluation receives an overall rating of 4.6 with an interpretation of Strongly barangay Poblacion 6 given to its residents

### V. CONCLUSION

The goal of this research is to provide an open data and geobased information system for barangays in the Philippines offering features to efficiently collect, analyze, and use residents, crime and health records generated in the community. Hence, this research is a ground research in building national information systems linking government agencies and services. The software provides features that can provide a set of guidelines and communication tools to ensure that information and government services are delivered. However, this research also has recommendations to enhance the findings of the study including: (a) incorporation of transactional component features on availing social security benefits, access to nearby hospitals and referrals to nearby agencies who could support other government related programs, (b) community should be educated about the open data systems through capacity building and effectiveness study, identify areas of improvement, and (c) integrate other open data requirements of the national government such as financial budget, existing projects handled by the barangay, and other government services to involve residents in planning, analysis of requirements, and development of relevant programs for the residents in the community.